\documentclass[aps,prl,twocolumn,superscriptaddress
,showpacs
]{revtex4}

\usepackage{amsfonts,amssymb,amsmath,latexsym,epsfig,wasysym} 
\usepackage[sort&compress]{natbib}

\usepackage{color,soul}

\begin{document}

\title{Noise-enhanced trapping in chaotic scattering \\ ({\small Phys. Rev. Lett. {\bf
      105}, 244102 (2010)})}

\author{Eduardo G. Altmann}
\affiliation{Instituto de F\'isica, Universidade Federal do Rio Grande do Sul, 91501-970 Porto Alegre, Brazil}
\affiliation{Max Planck Institute for the Physics of Complex Systems,  01187 Dresden, Germany}
\author{Antonio Endler}
\affiliation{Instituto de F\'isica, Universidade Federal do Rio Grande do Sul, 91501-970 Porto Alegre, Brazil}

\begin{abstract}
We show that noise enhances the trapping of trajectories in scattering systems. In fully chaotic systems,
the decay rate can decrease with increasing noise 
due to a generic mismatch between the 
noiseless escape rate and the value predicted by the Liouville measure of the exit 
set. In Hamiltonian systems with mixed phase space we show that noise leads to a slower
algebraic decay due to trajectories performing a random walk inside Kolmogorov-Arnold-Moser islands. 
 We argue that these noise-enhanced trapping mechanisms exist in most scattering systems and are
 likely to be dominant for small noise intensities, which is confirmed through a detailed
 investigation in the H\'enon map. 
Our results can be tested in fluid experiments, affect the fractal Weyl's law of
  quantum systems, and modify the estimations of chemical reaction rates based on phase-space
  transition state theory.
\end{abstract} 


\maketitle


The responses of nonlinear dynamical systems to small uncorrelated random perturbations (noise)
can be surprising and seemingly contradictory. Noise usually destroys fine structures of
deterministic dynamics, e.g. it fattens fractals~\cite{grassberger,ott,tel.open}, but
it can also combine constructively with the nonlinearities and increase the order of the 
system~\cite{noise.positive}. In chaotic {\em Hamiltonian} systems, investigations focused on the
effect of noise on anomalous transport~\cite{floriani,altmann.3} and, very
recently, on chaotic scattering~\cite{mills,seoane,rodrigues}.    

Chaotic scattering is a basic process of Hamiltonian dynamics~\cite{ott,gaspard,TG},
with fundamental applications in classical~\cite{sommerer} and
quantum~\cite{ott,fractalweyl} systems, and recent applications     
ranging from plankton populations~\cite{tel.open} to blood
flows~\cite{schelin} and even the origin of life~\cite{scheuring}.
Chemical  (dissociative) reactions of simple molecules are also scattering 
  processes where chaos is essential in the microcanonical phase-space formulation of transition state
  theory (TST)~\cite{pstst}.
Scattered trajectories perform transiently chaotic motion while trapped by
fine structures of the phase space, such as fractal nonattracting sets
and chains of Kolmogorov-Arnold-Moser (KAM) islands~\cite{ott,gaspard}. Noise
destroys the small scales of these structures~\cite{mills}, modifies the temporal decay of trajectories
in time from algebraic to exponential~\cite{seoane,rodrigues} with an exponent that
increases with noise~\cite{seoane}, and create otherwise forbidden escape paths~\cite{rodrigues}.
All these effects weaken the deterministic trapping.

In this Letter we show that noise also plays a
constructive role in chaotic scattering, 
enhancing the trapping of trajectories. First we introduce and scrutinize two different
mechanisms responsible for this surprising effect, arguing that they exist in very 
general circumstances.
 The first mechanism acts in fully chaotic systems and reduces the escape
rate of particles by blurring the natural measure of the system.
The second mechanism acts on mixed-phase-space systems and enhances trapping by throwing 
trajectories inside KAM islands. We confirm the generality of these
mechanisms through simulations in the conservative H\'enon map, and we explore the
  implications of our results to physical systems.

To illustrate how noise enhances trapping in fully-chaotic systems, consider the
baker map~\cite{ott} 
\begin{equation*}\label{eq.baker}
M:\left\{\begin{array}{ll}
(x_{t+1},y_{t+1})=(x_t/2,2y_t), & y\leq 1/2, \\
(x_{t+1},y_{t+1})=((x_t+1)/2,2y_t-1), &  y> 1/2,
\end{array}
\right.
\end{equation*}
defined in~$[0,1] \times [0,1]$. Escape is introduced through an arbitrary leak~$I$~\cite{paar,schneider,altmanntel,bunimovich}:
\begin{equation}\label{eq.leak}
\tilde{M}(x_t,y_t) = \left \{ \begin{array}{ll} M(x_t,y_t) & \text{ if } (x_t,y_t) \notin I \\
                                       \text{ escape } & \text{ if }  (x_t,y_t) \in I .\\
                                       \end{array} \right.
\end{equation}
For concreteness, consider~$I$ to be a vertical
stripe~($I=[x_c-\Delta_x,x_c+\Delta_x] \times [0,1]$)  at the center of the
map ($x_c=0.5,\Delta_x=0.05$). 
The survival probability~$P(t)$ of typical initial conditions
decays asymptotically as~$P(t) \sim e^{- t/\tau}$~\cite{ott,gaspard,TG}. 
In chemical reactions~$P(t)$ corresponds to the reactant lifetime distribution, which
plays a central role in TST~\cite{pstst}. 
The trapping strength is quantified through the characteristic life time~$\tau$, which is the
reciprocal of the escape rate and is different from the mean escape
time~\cite{altmanntel}. It can be obtained from periodic orbits, e.g. by
calculating the leading root~$z^\dagger$ of the (truncated) polynomial approximation of the 
zeta function~\cite{gaspard},
\begin{equation}\label{eq.upos}
1/\zeta(z)=\prod_p (1-z^{t_p}/\Lambda_p), \; \text{ as } \; \tau_{\xi=0} = \ln z^\dagger,
\end{equation}
where the product is taken over all periodic orbits~$p$, which have period~$t_p$ and 
expanding eigenvalue~$\Lambda_p$, and  $\xi=0$ indicates absence of noise~\cite{dettmann}. 
We locate the orbits analytically so that Eq.~(\ref{eq.upos})
yields~$\tau_{\xi=0}$ with a higher precision than simulations.
For the baker map example above, all~$1990$ orbits up to
period~$18$ yield~$\tau_{\xi=0}=6.06 \pm 0.02$.

\begin{figure}[!ht]
\includegraphics[width=1\columnwidth]{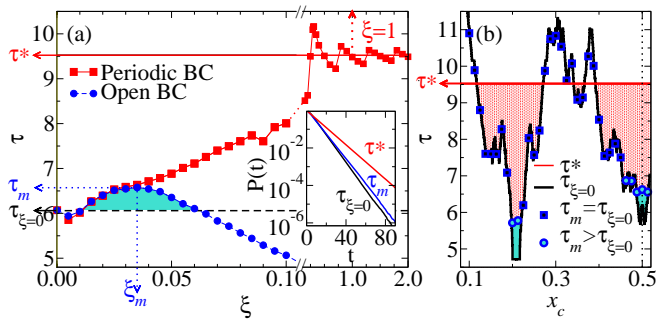}
\caption{ (Color online) Noise-enhanced trapping for fully chaotic systems. (a) The life time~$\tau$ for
  the baker map~(\ref{eq.leak}) is shown as a function of the noise
  intensity~$\xi$ for two BC: periodic (red~$\blacksquare$) and open (blue~$\newmoon$).  
  Horizontal lines: $\tau_{\xi=0}=6.06$ (solid) and~$\tau^*=9.49$ (dashed). 
Inset: $P(t)$ for~$\xi=0,\xi=\xi_m=0.035$ (open BC) and $\xi=1$ (periodic BC). (b) Values of~$\tau^*,\tau_{\xi=0}$ and $\tau_m$ for different positions~$x_c$
of the leak ($\Delta_x=0.05$). Noise-enhanced trapping
occurs when~$\tau^*>\tau_{\xi=0}$ for periodic BC and when~$\tau_m>\tau_{\xi=0}$
for open BC (blue shading).
}
\label{fig1}
\end{figure}

Next we consider the effect of noise, added independently to each trajectory $x\mapsto
x+\xi\delta_x, y\mapsto y+\xi\delta_y$, where $\xi$ 
controls the noise intensity and~$\delta_{x,y}\in[-1,1]$ are independent uniformly identically distributed
stochastic variables. Let us first consider the
simplest case of periodic boundary conditions (BC) for trajectories driven by noise outside~$[0,1]\times[0,1]$. This would be the natural
choice for maps defined on the torus.
In this case, 
for~$\xi\rightarrow\infty$ trajectories are uniformly distributed
in~$[0,1]\times[0,1]$ and $\tau\rightarrow\tau^*$ with~\cite{paar,schneider,altmanntel}
\begin{equation}\label{eq.gammastar}
\tau^*=[-\ln(1-\mu(I))]^{-1} \;[\approx 1/\mu(I) \text{ for small } \mu(I)],
\end{equation}
where $\mu(I)$ is the Liouville measure (phase-space area) of the leak~$I$. 
This is
  equivalent to the statistical microcanonical estimation of chemical reaction rates in  
TST~\cite{pstst}. In the baker's map
example above~$\mu(I)=2\Delta_x=0.1$ and~$\tau^*=9.49$ which is greater than the noiseless
case $\tau_{\xi=0}=6.06$ calculated above. The trapping is enhanced by noise! 

Let us take a closer look at the two crucial points that lead to this simple yet surprising
result. The first crucial point is the periodic BC that guarantees that~$\tau\rightarrow\tau^*$
for~$\xi\rightarrow\infty$. Another natural choice is open BC, in which case trajectories
outside~$[0,1] \times [0,1]$ escape. This type of escape prevails for large
$\xi$ and reduces~$\tau$ [for the baker map $\tau =-1/\ln((1-\Delta_x)/(2\xi)^2) \sim 1/\ln\xi$ for
$\xi\ge1$], but it is negligible for small~$\xi$. 
The 
results shown in Fig.~\ref{fig1}(a) confirm the existence of noise-enhanced trapping
for open BC, with a maximum~$\tau_m>\tau_{\xi=0}$ at~$\xi=\xi_m>0$. Noise
increases~$\tau$ by more than~$50\%$ in the periodic BC case and by almost~$10\%$ in
the open BC case.

\begin{figure}[!t]
\includegraphics[width=1\columnwidth]{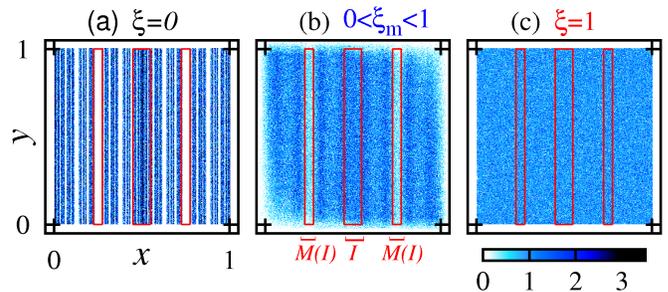}

\caption{(Color online) Noise tends to uniformly distribute surviving trajectories. Quasi-invariant
  density~$\rho_\xi$  of map~(\ref{eq.leak}) obtained at $t=40$ for: (a)
  $\xi=0$  ($\tau=\tau_{\xi=0}$), $\rho_{\xi=0}$ distributed along the unstable
  manifold of the chaotic saddle~\cite{ott,TG,gaspard}; (b) $\xi=\xi_m$ ($\tau=\tau_m$,
  open BC), $\rho_\xi$ is smoothed in~$I$ and negligible outside~$[0,1] \times [0,1]$; 
  (c) $\xi=1$ ($\tau=\tau^*$, periodic BC), $\rho_{\xi=1}=\rho_\mu$ is uniformly distributed. The
  leak~$I$ ($x_c=0.5,\Delta=0.05$) and its forward  iteration~$M(I)$ are indicated.} 
\label{figExtra}
\end{figure}

The second crucial point in the derivation above is~$\tau^*>\tau_{\xi=0}$. 
References~\cite{paar,schneider,altmanntel,bunimovich} show
that this holds for most leaks, also for map~(\ref{eq.leak})~\cite{schneider},
what is confirmed in Fig.~\ref{fig1}(b). For periodic BC the condition~$\tau^*>\tau_{\xi=0}$ is
sufficient for the existence of noise-enhanced trapping. For open BC the
condition~$\tau^*>\tau_{\xi=0}$ is necessary but not sufficient, and noise-enhanced trapping exists
only when~$\tau_m>\tau_{\xi=0}$.  In Fig.~\ref{fig1}(b) this is seen only in
the deep ``valleys'' of the~$\tau_{\xi=0}(x_c)$ landscape. But the noise enhanced {\em
  mechanism} is also present elsewhere, e.g. 
for several~$x_c$'s the~$\tau(\xi)$ curve has local maxima. 
The difference between~$\tau^*$ and~$\tau_{\xi=0}$ is a consequence of the difference
between the Liouville invariant density~$\rho_\mu$ (uniform in $x,y$) and  the open
system's quasi-invariant density~$\rho_{\xi=0}$ (nonuniform in $x,y$)
inside~$I$~\cite{paar,altmanntel}. Note that 
escape occurs one iteration {\em after} trajectories enter~$I$ (or
leave~$[0,1]\times[0,1]$). When~$\rho_{\xi=0}$ is large in~$I$, the open system measure
of~$I$ is larger than~$\mu(I)$ and $\tau^*>\tau_{\xi=0}$, see Eq.~(\ref{eq.gammastar}). In this case the
dominant effect of noise is to move trajectories outside~$I$, avoiding their escape and
increasing~$\tau$. More generally, noise acts on $x,y$ coordinates making~$\rho_\xi$ 
more uniform in~$x,y$~\cite{tel.open,mills}, departing from~$\rho_{\xi=0}$ and
approaching~$\rho_\mu$.  This $\tau$-increasing effect is shown in Fig.~\ref{figExtra}~\cite{paar,foot2}.

The mechanism described above acts on transient chaos and is different from stochastic and coherence
resonance~\cite{noise.positive}. It has been observed in
dissipative maps near crisis~\cite{franaszek,reimann} and on extended excitable
systems~\cite{spatiotemporal}. Strong nonlinearity was shown to be a sufficient condition
for its occurrence in~$1$-d maps~\cite{reimann}.
No effect was observed in a $2$D system~\cite{blackburn}, raising
doubts about its generality in higher dimensions. Here we show the existence of this mechanism in (Hamiltonian)
scattering systems. We identify~$\tau^*>\tau_{\xi=0}$ as the crucial condition,
clarifying the generality  also for higher dimensions. 
A similar approach in $1$D maps related this mechanism to the nonuniformity of the invariant
density of the closed system~\cite{faisst}. In Hamiltonian systems the closed system
density~$\rho_\mu$ is uniform, but we show that the mechanism is nevertheless effective because the open system density~$\rho_{\xi=0}$ is
non-uniform.

\begin{figure}[!ht]
\includegraphics[width=1\columnwidth]{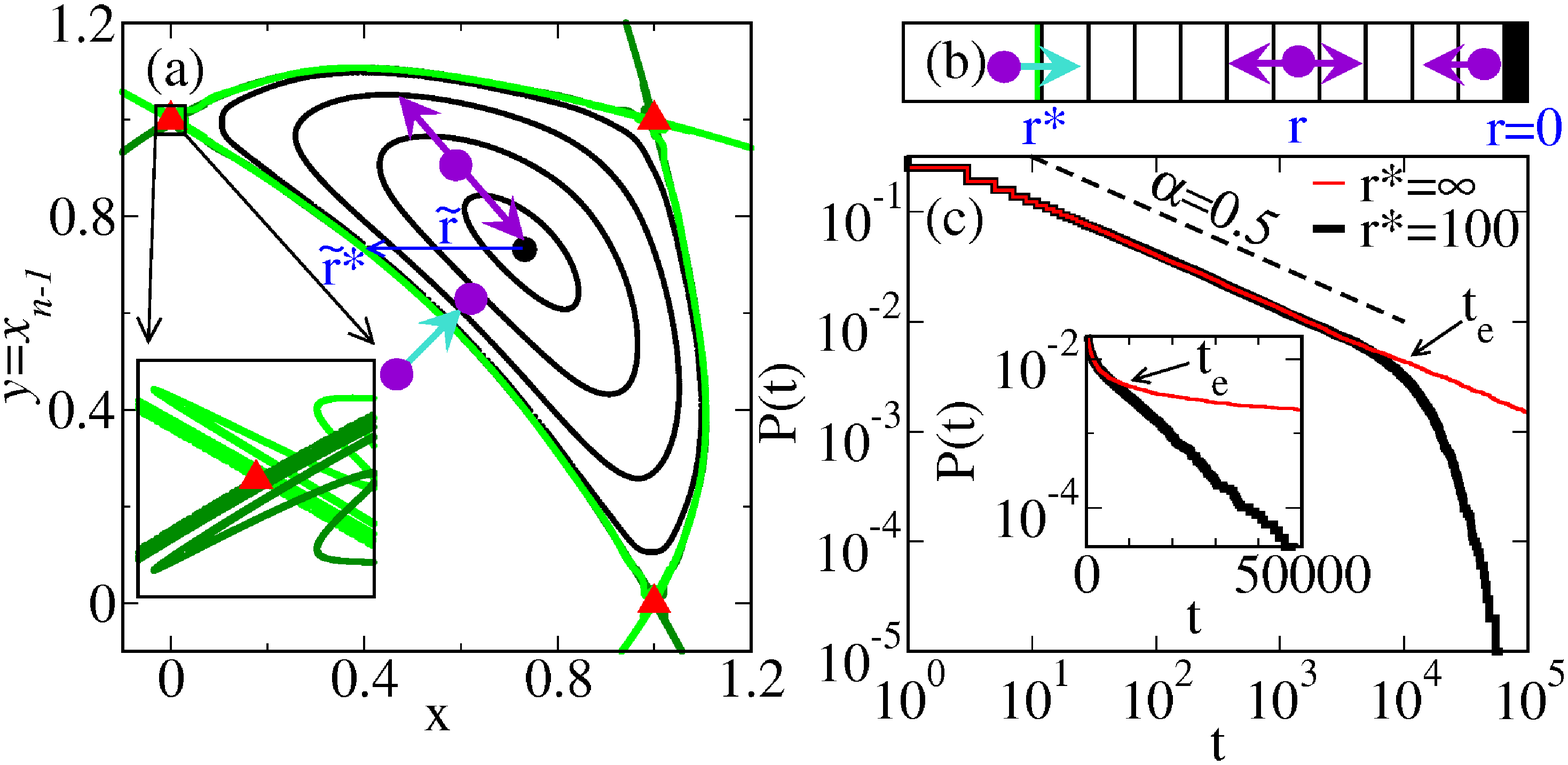}
\caption{(Color Online) Random-walk model for
trajectories inside KAM islands. (a) Illustrative KAM island [H\'enon map~(\ref{eq.henon})
with $k=2$]. Inset: Magnification around the period three orbit~$\blacktriangle$ showing
the intersection of manifolds. (b) Random-walk model with probabilities $p_\pm=0.5$ of 
stepping $\Delta r = \pm 1$,  one
reflecting ($r=0$, center of the island) and one absorbing boundary ($r=r^*\sim \tilde{r}^*/\xi$, where
$\tilde{r}^*$ is proportional to island width and $\xi$ to step length).
(c) $P(t)$ for a simulation of $10^5$ walkers started in~$r=r^*-1$, with $r^*=\infty$ (thin red 
line) and~$r^*=100$ (thick black line). Dashed line: Scaling $\alpha =0.5$. Inset: Linear-log plot emphasizing
the exponential tail (for~$r^*=100$) starting at~$t_e \approx 10^4=(r^*)^2\sim (\tilde{r}^*/\xi)^2$.}
\label{fig2}
\end{figure}

In mixed-phase-space systems the exponential decay discussed so far is replaced by an 
algebraic decay~$P(t) \sim t^{-\alpha}$ due to the stickiness of 
KAM islands~\cite{altmann.3,zaslavsky}.
Figure~\ref{fig2} illustrates how noise enhances trapping in this case. If~$\xi$ is larger
than the lobes of manifold intersections close to the KAM island (see inset), it
will be more likely for trajectories to approach (or enter) the region corresponding to
the KAM island in the deterministic dynamics by jumping over the manifolds that shield the
island. The time scales of this effect can be estimated following
Ref.~\cite{floriani}, wich argues that for small~$\xi$ the deterministic algebraic decay
will be interrupted at a time
%
$t_b \sim 1/\xi^\beta$,
%
where $\beta\approx1$ depends on~$\alpha$. 
An asymptotic exponential decay after such cut-off is the typical effect of noise in intermittent
systems~\cite{floriani}, observed also in scattering
systems~\cite{seoane}. Our approach differs from
Refs.~\cite{floriani,seoane} because we consider an additional trapping regime
dominated by trajectories that jump inside the island, circle  the elliptic fixed point, and perform a
random walk in the perpendicular direction. This $1$D random-walk model  can be solved
analytically in infinite domain~\cite{feller} and yields~$P(t)\sim t^{-\alpha}$
with~$\alpha=0.5$. This is smaller than 
the deterministic 
exponent~$1<\alpha_{\xi=0}<2$, meaning that trapping is enhanced. In
reality, the island has a finite area that corresponds to a reflecting boundary in the
model, as 
explained in Fig.~\ref{fig2}(b). This introduces a cutoff~$t_e>t_b$ in
the~$\alpha=0.5$ algebraic decay followed by an exponential decay, as shown in
Fig.~\ref{fig2}(c). This second cutoff occurs when
trajectories explore the full island of size~$\tilde{r}^*$, and since this is a diffusive
process~$\tilde{r}^*=r^*/\xi \sim \sqrt{t}$ and $t_e\sim 1/\xi^2$~\cite{altmann.3}.
For~$\xi\rightarrow0$ the interval of enhanced trapping~$\Delta t \equiv (t_e-t_b)
\rightarrow \infty$ because of the different scalings of~$t_b$ and~$t_e$.
Therefore, even if not valid asymptotically, enhanced
trapping is in practice dominant for small~$\xi$. Similar models have been used to investigate
different problems: anomalous 
 transport~\cite{altmann.3} and escape of initial conditions inside KAM islands in
 random maps~\cite{rodrigues}.

\begin{figure}[!bt]
\includegraphics[width=1\columnwidth]{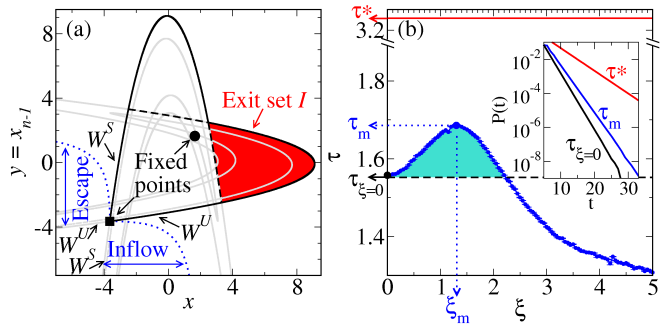}
\caption{ (Color online) Noise-enhanced trapping in the fully chaotic $(k=6)$ H\'enon
  map~(\ref{eq.henon}). (a) Phase space with trapped region, delimited by the solid
  black line (first intersection of the manifolds $W^{U,S}$ of the fixed point~$\blacksquare$). Escape
  (Inflow) denotes an unbounded forward (backward) invariant set that
  diverges in forward (backward) iterations~\cite{petrisor}.  
Initial conditions and the region where trajectories are removed were chosen inside these sets~\cite{foot1}. 
The exit set~$I$ is the set of points that exit the trapped region in one iteration~\cite{meiss}.
(b) Dependence of $\tau$ on~$\xi$ with $\tau_{\xi=0}=1.557$ (dashed line)
and~$\tau^*=3.239$ (dotted line). Inset: $P(t)$ for~$\xi=0,\xi=\xi_m,$ and with~$\tau^*$.}
\label{fig3}
\end{figure}

\begin{figure}[!bt]
\includegraphics[width=1\columnwidth]{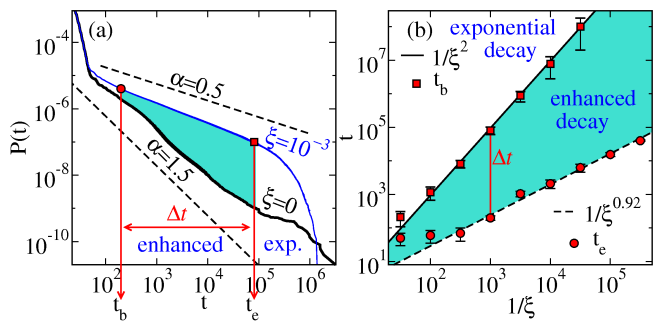}
\caption{ (Color online) Noise-enhanced trapping in the H\'enon map~(\ref{eq.henon}) with
  mixed phase space ($k=2$). (a) $P(t)$ for~$\xi=0$
  (thick black line) and~$\xi=10^{-3}$ (thin blue line)~\cite{foot1}. The times~$t_b$ and~$t_e$ indicate the
  beginning and end of the mechanism described in Fig.~\ref{fig2}. Power-law scalings are shown as a reference (dashed lines). 
(b) Dependence of~$t_b$ and~$t_e$ on~$\xi$ with fits of the model predictions
(with $\beta=0.92$; see text).
}
\label{fig4}
\end{figure}

Finally, to confirm the general validity of the two mechanisms discussed above, we
investigate the paradigmatic H\'enon map~\cite{meiss,petrisor,endler} 
\begin{equation}\label{eq.henon}
x_{t+1}=k-x^2_t-x_{t-1}+\xi \delta_t,
\end{equation}
where different random perturbations~$\delta_t$, defined as before, are applied to each trajectory.
To investigate the mechanism for fully chaotic systems we
choose~$k=6$ in Eq.~(\ref{eq.henon}), which is the smallest integer for which a complete
horseshoe exists~\cite{endler}. 
In Fig.~\ref{fig3}(a)
we identify in the phase space of
map~(\ref{eq.henon}) the trapped region and exit set~$I$
(reactant region and transition state in TST~\cite{pstst}). 
We estimate graphically~\cite{meiss} $\mu(I)=\text{area}(I)/\text{area(trapped region)}=0.2656$, which leads 
through Eq.~(\ref{eq.gammastar}) to~$\tau^*=3.239$. We compute analytically~\cite{endler}
all $226$ periodic orbits up to period~$10$ and
through Eq.~(\ref{eq.upos}) we obtain $\tau_{\xi=0}=1.557
\pm0.001$. Again~$\tau^*>\tau_{\xi=0}$,
fulfilling the necessary condition identified above for the existence of noise-enhanced
trapping. Our numerical simulations reported in Fig.~\ref{fig3}(b) confirm the
nonmonotonic dependence of~$\tau$ on~$\xi$ with~$\tau_m>\tau_{\xi=0}$, in
perfect analogy with the results for the baker map shown in Fig.~\ref{fig1}. 
To investigate the mechanism for mixed phase-space systems we choose~$k=2$ in
Eq.~(\ref{eq.henon}). The phase space is similar to the one
shown in Fig.~\ref{fig3}(a) except for the KAM island in Fig.~\ref{fig2}(a) around the
fixed point (black dot~$\newmoon$). The results are summarized in Fig.~\ref{fig4} and confirm quantitatively
 the predictions 
(scalings of~$t_{b,e}$ and~$\alpha=0.5$) of the random-walk
model in Fig.~\ref{fig2}(b,c).

In summary, we have shown that weak noise leads to a slower decay of the survival
probability~$P(t)$ in fully chaotic and in mixed-phase-space scattering systems. 
In both cases $P(t)$ changes nonmonotonically with noise intensity~$\xi$
(Figs.~\ref{fig3} and~\ref{fig4}), an effect previously observed in the dependence of
the diffusion coefficient on~$\xi$~\cite{altmann.3,klages}. 
Our approach to the mechanism for fully chaotic systems extends naturally to higher dimensions, while  the mixed-phase-space mechanism has to be
expanded to take Arnold's diffusion into account~\cite{ott}.  

All scattering systems are likely to be subject to small noiselike perturbations and
to experience an enhancement of the trapping.
Experiments on the chaotic advection
of passive tracers~\cite{sommerer} can provide a direct test of our predictions:
by systematically changing the properties of the tracer and fluid 
one can control the molecular diffusion ($\sim\xi^2$)
and obtain the dependence of~$P(t)$ on~$\xi$. 

Our results also describe the effect of noise in the phase-space formulation of TST, where
an increase of the chemical reaction rate is expected.
Simulations of unimolecular reactions with mixed phase space found $P(t)~t^{-\alpha}$
with~$\alpha<1$ and an exponential
  tail. According to Ref.~\cite{pstst} the dynamical origins of this behavior remain
  obscure. We show that noise leads to the same observations. It remains to be shown
  whether this explains the previous simulations, where noiselike perturbations could originate from
  round-off errors or from higher degrees of freedom (see Ref.~\cite{altmann.3}).
Our results also impact quantum systems. For instance, the increase in the 
characteristic life time~$\tau$ indicates a modification of the fractal dimensions of the
invariant sets~\cite{romeiras,tel.open,foot2} and consequently of the
fractal Weyl's law~\cite{fractalweyl}.

\acknowledgments
We are indebted to T. T\'el and D. Paz\'o for insightful suggestions. E. G. A. was supported by an Max Planck Society Otto Hahn fellowship.



\begin{thebibliography}{99}
\bibitem{foot2} We have verified that trapping is enhanced also for random
  maps, where 
dimensions remain fractal~\cite{romeiras,tel.open}.
\bibitem{foot1} Trajectories were started at $x_i\in \{0.9
  x_{fp},1.0 x_{fp}\},y_i=1.1 y_{fp}$ and removed at~$x<1.1 x_{fp},y<1.1 y_{fp}$,
  where $x_{fp}=y_{fp}=-1-\sqrt{1+k}$~\cite{meiss,petrisor,endler}.
\end{thebibliography}
\end{document}